# Active nitrogen flux measurement during GaN growth based on the transmitted signal detected with a pyrometer

**Matteo Canciani, Stefano Vichi, Oksana Koplak, Sergio Bietti and Stefano Sanguinetti**

University of Milan-Bicocca, via R. Cozzi, 55, 20126, Milan (Italy)

E-mail: stefano.vichi@unimib.it

**Abstract.** A novel approach for the measurement of the Nitrogen active species generated by a plasma source in the molecular beam epitaxy environment is here presented. The method is based on the analysis of the variations in the optical signal measured by a pyrometer during a two step, Gallium rich and Nitrogen controlled, growth modes. The method permits a precise, quantitative and direct measurement of the flux of active species as a function of the plasma generation parameters of the cell: nitrogen gas flux and RF-power.

4# 1. Introduction

Compound semiconductors of the III-N family have a great interest for their advantage for devices fabrication as light emitting diodes (LEDs) [1], laser diodes (LDs) [1], high-electron mobility transistors (HEMTs) [2] and biosensors [3]. Specifically, the advantages of Nitrides semiconductors, like GaN, comprehends higher breakdown strength, faster switching speed, higher thermal conductivity, lower on-resistance and thermal stability [4, 5].

Among the various techniques available for the synthesis of III-nitride semiconductors based devices, Plasma-Assisted Molecular Beam Epitaxy (PA-MBE) stands out as a highly promising method [6, 7]. This technique offers unique advantages, particularly for the growth of high-In concentration $In_xGa_{1-x}N$ ternary alloys, which are of significant interest for applications in fields such as sensors and photovoltaics. Unlike Metal Organic Chemical Vapor Deposition (MOCVD) and $NH_3$-Assisted MBE, PA-MBE does not require the thermal decomposition of $NH_3$ to supply reactive nitrogen during growth [8]. This key feature enables growth at lower substrate temperatures, making possible to synthesize alloys such as $In_xGa_{1-x}N$, which are otherwise inclined to thermal decomposition. Furthermore, PA-MBE overcome issues related to hydrogen passivation, offering additional benefits for achieving high-quality materials and optimizing their electronic and optical properties [9].

The precise evaluation of the source beams in terms of flux of active species that is directed towards the substrate is of outstanding importance in all the plasma assisted growth processes. In the field of III-nitride semiconductor materials, the case of the calibration of the nitrogen plasma flux is rather complex [10], as the usual calibration in terms of Beam Equivalent Pressure (BEP) using a pressure gauge positioned in the substrate position cannot be performed without the gauge degradation. Furthermore, the situation is complicated by the presence of different species in the plasma beam: molecules in the ground state ($N_2$), neutral atoms (N), excited radicals ($N_2^*$), molecular ions ($N^{2+}$) and atomic nitrogen ions ($N^+$) [11, 12, 13].

Several methods for the nitrogen plasma flux detection and the related GaN growth rate can be found in the literature, like RHEED signal monitoring during GaN deposition [14, 15, 16, 17], reflectivity measurements [18], infrared (IR) radiation interferometry using a fiber coupled optical pyrometer [19] and spectroscopic ellipsometry [20]. Although these are all valid techniques, each presents certain limitations and complexities in terms of application and implementation. For instance, Reflection high-energy electron diffraction (RHEED) signal monitoring relies on detecting signal variations during the deposition process, which requires precise focusing of the RHEED beam to obtain accurate and reproducible data, as well as addressing potential local surface heating due to the high-energy electron beam incidence. However, the implementation of a reflectivity setup requires the aid of extra components outside the MBE apparatus, meticulous alignment in a vibration-free environment, and, most importantly, the availability of two symmetrical viewports relative to the sample position in the growth chamber.

Here, we present a different approach which permits to determine the active nitrogen flux responsible for the GaN growth in a PAMBE environment on the basis of the signal variations read by a pyrometer pointing towards the sample's surface. This method does not require any additional external equipment beyond the pyrometer, which in an MBE environment is an essential tool for the precise detection of the substrate surface temperature during the whole growth process. Furthermore, being an optically based detection technique, it does not need ultra-high vacuum (UHV) to be performed. Therefore, this method is also suitable for use in Metal Organic Chemical Vapor Deposition (MOCVD) systems as well.

# 2. Experimental

There are several sources of nitrogen plasma that can be found in the market, such as Electron Cyclotron Resonance (ECR) and ammonia surface cracking ($NH_3$). This article uses the activation of an inert nitrogen gas by radiofrequency (RF) using a high-density radical source (HDRS) from Hakuto. The experiments were performed using an MBE system equipped with a standard Knudsen cell for gallium (Ga) and a HDRS for nitrogen plasma generation. The plasma source uses high-frequency electromagnetic waves (EM) of 13.56 MHz to ionize the nitrogen gas for plasma production. The two main parameters of this cell are the flux of inert nitrogen gas supply in terms of standard cubic centimeters per minute (sccm) and the radiofrequency power (RF-power) of the EM field, which can be tuned from 0 to 600W, with a precision of 1W.

The MBE set-up was equipped with an Infrared (IR) Photrix pyrometer with a spectral response peaked at 900 nm and a detection spot of approximately $\sim$ 5 mm in diameter, and a conventional RHEED set-up.

The substrates used for the depositions were commercial undoped GaN/Sapphire (0001) templates. To improve heat absorption, on the back side of the templates a $\sim$ 50 nm Ti layer was deposited by e-beam evaporation.

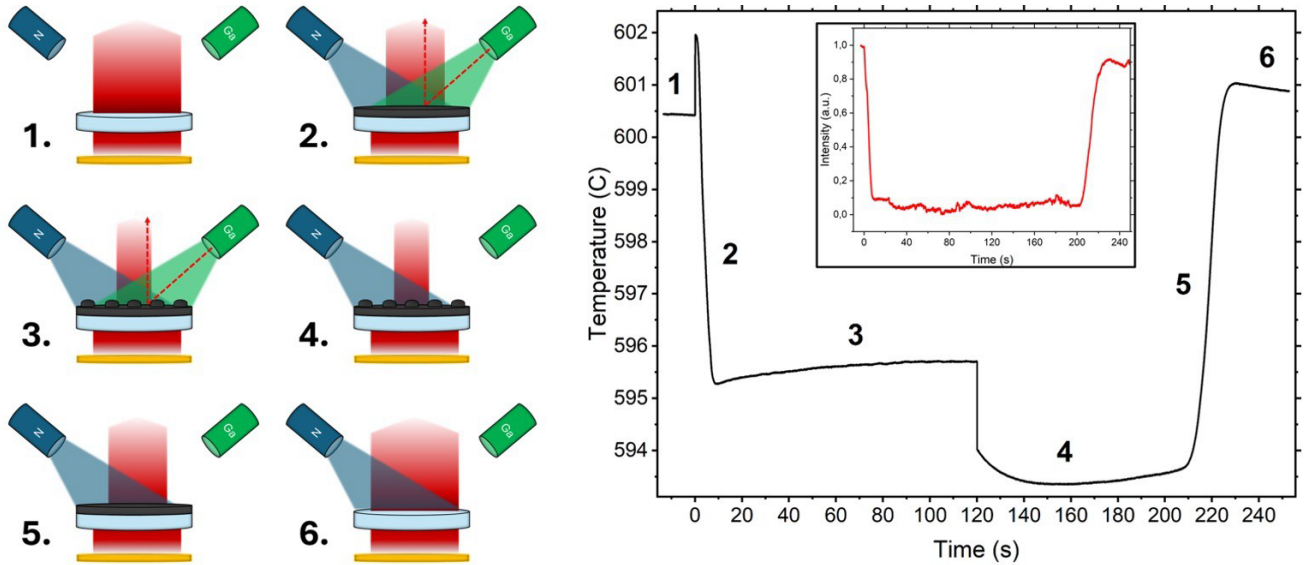

**Figure 1.** Left panel: schematic of the entire process of GaN deposition and residual Ga crystallization used for the calibration. The size of the red arrow pointing upwards is proportional to the expected emissivity of the GaN surface on different conditions. 1) steady state condition without Ga and N fluxes, 2) initial Ga monolayers accumulation during the GaN growth, 3) Ga droplets formation, 4) Ga droplets consumption after the Ga flux has been switched off, 5) residual Ga monolayers crystallization, 6) restoration of the initial steady state. Right panel: plot of signal detected by the pyrometer during the process schematized in the left panel. The different steps of the process, from 1 to 6, are indicated. The inset panel shows the same process monitored with RHEED.

The substrates were subjected to a thermal treatment to prepare the surface for epitaxy. First, a 2 hour annealing step was performed at 200 °C in the intro chamber at a pressure of $1 \times 10^{-8}$ Torr, followed by an additional 18 h of annealing in the main chamber, again at 200 °C, with a base pressure of $5 \times 10^{-10}$ Torr. Before growth, all samples were heated to 750 °C for 30 min to remove native surface oxide and the in situ dynamic of the process was monitored by RHEED until a clear 2D pattern was present.

Each measurement was acquired following a two-step scheme: a) GaN deposition for 120 seconds under Ga-rich conditions that promote metallic Ga accumulation on the growing surface, b) closure of the Ga shutter while keeping the nitrogen plasma active to allow residual Ga crystallization; see Figure 1. During the deposition processes, both the RHEED and the pyrometer signals were monitored and analyzed in real time. All data were acquired with a constant Ga flux of $1.1 \times 10^{-7}$ Torr and a substrate temperature of 600 °C to prevent Ga desorption from the surface [21, 22]. To improve the accuracy of the measurements, the substrate rotation was stopped during depositions and the substrate heater power was kept fixed. The nitrogen plasma parameters were varied from 2 to 5 sccm and from 300 to 500 W. After each series, the surface was smoothed by growing few nanometers thick GaN layer at high temperature to recover the 2D surface, as confirmed by the 2D RHEED pattern, to ensure all measurements began under the same conditions.

## 3. Results and discussion

The right panel of Figure 1 shows the typical signal acquired by the pyrometer during the deposition of GaN in the Ga-rich regime and the subsequent residual crystallization of Ga. In the specific case of the right side of Figure 1, the signal was obtained with plasma parameters of 300 W and 3 sccm. All notable transients are marked on the signal and are correlated with different surface processes. The inset of Figure 1 shows the intensity variation of the specular RHEED spot during the same deposition, which allowed us to confirm the nature of these surface processes compared to previous works in the literature [23, 24]. The RHEED signal is noticeably noisier than the pyrometer signal, which displays distinct specific behavior that facilitates a clearer distinction between the surface processes occurring.

As illustrated in the left side of Figure 1, the pyrometer signal can be divided into six distinct regions. 1) *Steady state* condition, where no material is deposited. The pyrometer signal measures the temperature of the titanium (Ti) back coating, as the GaN is transparent at the pyrometer detection wavelength. 2) *Deposition* of both Ga and N under Ga-rich conditions. The pyrometer signal drops due to the accumulation of excess Ga on the surface, forming a metallic layer [16, 23], which reduces the transmission of radiation emitted by the Ti coating reaching the pyrometer. Before this drop, a sharp rise is observed, attributed to radiation emitted by the open Ga cell and reflected from the sample surface. 3) After a critical thickness of the metallic layer is reached, the signal stabilizes showing a *plateau*. The slight increase in the signal is attributed to the enhanced reflected radiation of the Ga cell, which increases because of the accumulation of metals on the surface. At the end of this step, as can be seen by the abrupt drop of the signal, similar to the steep increase between steps 1-2, the Ga cell is closed while the nitrogen continues to be supplied. 4) The Ga reservoir accumulated during step 3 is being consumed by the nitrogen flux via *GaN crystallization*. Similarly to the previous step, the signal remains almost flat, with smaller deviations attributed to spurious background radiation. 5) *Steep increase* in the pyrometer signal, caused by the crystallization of the residual metallic layer. 6) restoration of *steady condition*, with all cells closed. The slightly higher temperature compared to step 1 is due to a drift of the surface temperature due to the choice of keeping the heater power constant throughout the process.

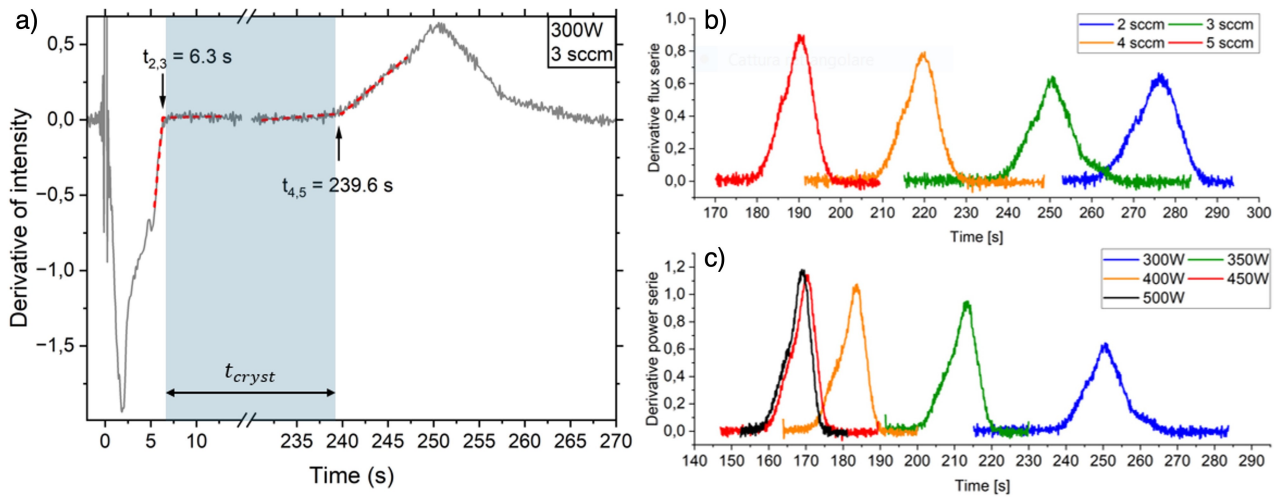

**Figure 2.** a) Derivative of the pyrometer signal measured with a nitrogen plasma flux of 3 sccm and a plasma source power of 300 W. The derivative is used to improve the signal analysis and exhibits clear indications on the signal intensity variations. The two important time transients used for our calibration are highlighted in this plot, showing a transition between steps 2 and 3 at $t_{2,3} = 6.3\ s$ (Ga monolayers accumulation and Ga droplet formation) and a transition between steps 4 and 5 at $t = 239.6\ s$ (Ga droplet consumption and residual Ga monolayers crystallization). b) and c) derivative of the pyrometer signals of the sccm and power series, respectively, showing the clear peaks dependence.

It is crucial to understand the meaning of the variation in the pyrometer signal during deposition and its relation to kinetic surface processes, which is closely related to the combination of the emissivity of the GaN surface and of the Ti back coating. Under steady conditions, the pyrometer detects the temperature of the Ti back coating, as GaN is transparent at 900 nm. During depositions, this signal is reduced due to the accumulation of metallic Ga on the surface [?], which forms a reflective layer that acts as a mirror for a specific amount of black-body radiation coming from the Ti coating at the back of the substrate and is not compensated for the emissivity of the thin Ga layer that accumulated at the GaN surface. In addition, also the radiation emitted by the open Ga cell and reflected by the sample contributes to the signal read by the pyrometer in the form of a constant background, as can be seen in the steep increase (decrease) in the signal between steps 1 and 2 (steps 3 and 4) in Figure 1.

It is important to emphasize the extreme precision in the detection of surface process variations of this optical technique. This high sensitivity is achieved because of the high reflectivity of the metals, which significantly reduces the intensity of the transmitted radiation.

After the formation of flat layers, Ga droplets begin to accumulate on the surface [16], although their effect on transmitted radiation remains negligible as long as they cover the surface partially.

When the growth rate is not too fast, growth proceeds at the GaN-Ga interface, gradually depleting the Ga droplets without crystallizing them. This behavior allows the formation and consumption of the droplets reservoir to serve as a relative calibration for Ga and nitrogen plasma fluxes.

Because the surface temperature in our experiments is low enough to avoid Ga desorption, a simple mass conservation model can be used to describe the system:

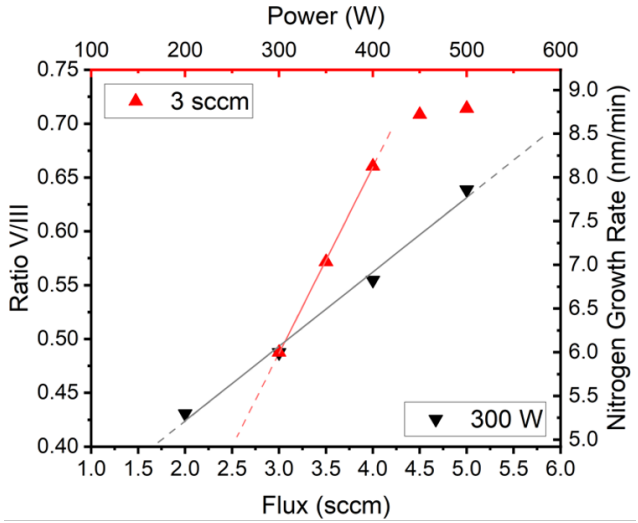

**Figure 3.** Results of both the nitrogen flux and RF-power series. The left and right axis respectively show the V/III ratio and the nitrogen growth rate (nm/min) for the nitrogen flux serie (black triangles) and power serie (red triangles). The linear behaviour is highligthed in the plot for both the graphs. It is worth mentioning the approach to the saturation self-limiting condition in the RF-plasma serie.

$$(\phi_{Ga} - \phi_N)t_{dep} = \phi_N\, t_{cryst} \tag{1}$$

where $t_{dep}$ is the time during which both Ga and N are deposited (that is, steps 2 and 3), and $t_{cryst}$ is the time required to crystallize the residual Ga (that is, steps 4 and 5). Here, $\phi_{Ga}$ and $\phi_N$ represent the Ga and nitrogen plasma fluxes, respectively. Since the amount of excess Ga deposited in step 2 is the same as the one crystallized in step 5, these terms can be canceled from the equations, yielding the following equation to determine the V/III ratio:

$$\frac{\phi_{N_2}}{\phi_{Ga}} = \frac{t_3}{t_3 + t_4} \tag{2}$$

To determine times $t_3 = (t_{dep})$ and $t_4 = (t_{cryst})$, it is necessary to precisely identify the transitions between steps 2 and 3, and steps 4 and 5. To improve the precision of the determination of the time transients, it is worth plotting the first derivative of the pyrometer signal with a broken linear fit function around the transition points, as shown in the left side of Figure 2. In Figure 3 the results of this analysis are shown. The left axis shows the V/III ratio (following Eq. 2) as a function of nitrogen flux with a fixed RF-plasma source power of 300 W (black triangles) and as a function of RF-plasma source power with a fixed nitrogen flux of 3 sccm (red triangles). On the right axis, the same data are shown in the correspondence of the equivalent GaN growth rate in the nitrogen-limited growth mode. A well-defined linear behavior of the nitrogen-limited growth rate and V/III ratio is obtained from each series (flux and power), as shown with the black and red linear fit, respectively, of Figure 3. It is important to notice that a saturation condition is obtained in the RF-Power series for values above 400-450 W, suggesting the reaching of the self-limiting condition of the nitrogen plasma crystallization at a constant flux of 3 sccm. From an accurate analysis of the linear behavior in Figure 3, the different contribution of the power and nitrogen gas flux to the nitrogen plasma generated flux can be highlighted. For instance, the same growth rate (or the equivalent V/III ratio) can be obtained both with (300W, 4sccm) and (340W, 3sccm) in the parameter space. This outcome is of outstanding importance in an PA-MBE environment, in which the background pressure during growth is a fundamental parameter and allows for a conservative choice of parameter (340W, 3sccm) to lower the pressure in the growth chamber during plasma growths.

As a matter of fact, these data allow us to control the nitrogen plasma flux and, following Eq. 2, to provide a quantitative measurement of it. All the data can be obtained from in situ measurements of the pyrometer intensity signal variations, thus permitting a clear and simple calibration of the nitrogen flux.

Figure 4 shows the time evolution of the pyrometer signal during Ga-rich growth under a nitrogen flux of 3 sccm and a plasma source power of 300 W. A clear interference pattern is visible, with an oscillation period $T = \lambda \backslash n \cdot cos(\vartheta) \cdot \rho$, where n is the GaN refractive index, $\vartheta$ is the angle between the pyrometer and the normal to the substrate, and $\rho$ is the growth rate [25, 26, 27]. In our specific case, we obtained a GaN growth rate $\rho$ of 11.92 ± 0.01 nm/min, assuming an error of $\lambda \pm 1$ nm and an error of ±0.2 s in $T$, as obtained by fitting the curve of Figure 4. Because the growth was performed under the nitrogen-limited regime, it is possible to directly determine that the equivalent nitrogen growth rate is half that of GaN. Finally, its linear relationship with the nitrogen flux calculated from equation 2 for a fixed Ga flux, shown in Figure 3, allows us to determine the equivalent nitrogen growth rate for all the parameters under study.

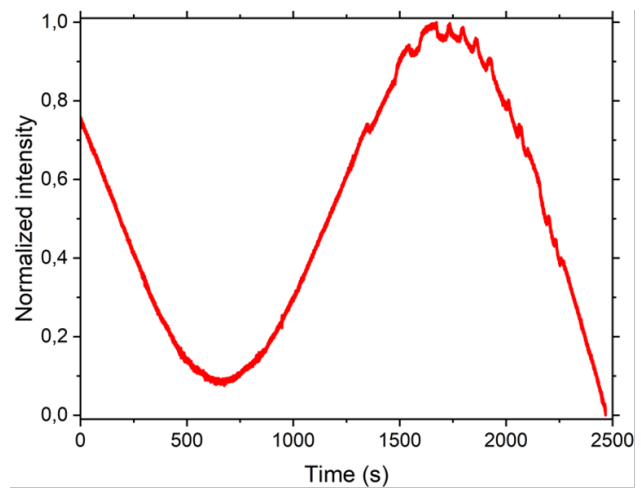

**Figure 4.** Time evolution of the pyrometer signal during a slightly Ga-rich GaN growth. This thin film interference is used here to gain information about the growth rate in N-limited growths. The nitrogen plasma generation parameters where 3 sccm and 300 W of RF-power.

This method provides an extremely precise calibration of all growth rates without the need of RHEED or symmetric MBE ports as needed for a reflectometry set-up, as demonstrated by the desorption measurements.

## 4. Conclusions

In this study a novel approach for the measurement of nitrogen plasma flux during GaN growth in terms of BEP was presented on the basis of a two-step procedure. The relationship between the intensity of the black body signal acquired by a IR Photrix pyrometer during all the acquisitions was correlated with the adatom kinetics on the growing GaN surface, allowing for the discrimination between metallic Ga accumulation and its GaN crystallization using the nitrogen plasma supply. This allows for the precise quantification of the amount of nitrogen plasma for both its generation parameters: nitrogen gas flux and RF-power of the cell. A well-established linear behavior is obtained in both parameters series. In the particular case of the RF-power series the saturation condition is reached, showing the generation self-limiting condition of the plasma flux for RF-power higher than 450W for a nitrogen gas flux of 3sccm. The use of the pyrometer for such analysis allows to a better reproducibility of the acquired signal when compared to other in-situ techniques as RHEED and allows for a 2D flattening of the GaN surface after each acquisition series.

## Acknowledgments

This work was funded by the National Plan for NRRP Complementary Investments (PNC, established with the decree-law of 6 May 2021, n. 59, converted by law n. 101 of 2021) in the call for the funding of research initiatives for technologies and innovative trajectories in the health and care sectors (Directorial Decree n. 931 of 06-06-2022) - project n. PNC0000003 - AdvaNced Technologies for Human-centrEd Medicine (project acronym: ANTHEM). This work reflects only the authors' views and opinions, neither the Ministry for University and Research nor the European Commission can be considered responsible for them.